\documentclass[review,
number,sort&compress,12pt]{elsarticle}

\usepackage{graphicx}
\usepackage{amssymb}
\usepackage{lineno}

\begin{document}
\begin{frontmatter}

\title{Calibration of Photomultiplier Tubes
for the Fluorescence Detector of Telescope Array Experiment
using a Rayleigh Scattered Laser Beam}

\author[saidai]{Shingo~Kawana\corref{SK}}
\author[ocu]{Nobuyuki~Sakurai}
\author[ocu]{Toshihiro~Fujii}
\author[icrr,ipmu]{Masaki~Fukushima}
\author[saidai]{Naoya~Inoue}
\author[utah]{John~N.~Matthews}
\author[ocu]{Shoichi~Ogio}
\author[icrr]{Hiroyuki~Sagawa}
\author[icrr,eri]{Akimichi~Taketa}
\author[icrr]{Masato~Takita}
\author[utah]{Stan~B.~Thomas}
\author[tit,tit2]{Hisao~Tokuno}
\author[tit]{Yoshiki~Tsunesada}
\author[kanagawa]{Shigeharu~Udo}
\author[colorado]{Lawrence~R.~Wiencke}

\cortext[SK]{Corresponding author.
E-mail address: kawana@crsgm1.crinoue.phy.saitama-u.ac.jp (S.~Kawana)}

\address[saidai]{Graduate School of Science and Engineering, 
Saitama University, Saitama 338-8570, Japan}

\address[ocu]{Graduate School of Science, Osaka City University, 
Sumiyoshi, Osaka 558-8585, Japan}

\address[icrr]{Institute for Cosmic Ray Research, University of Tokyo, 
Kashiwa, Chiba 277-8582, Japan}

\address[ipmu]{Institute for the Physics and Mathematics of 
the Universe, University of Tokyo, 
Kashiwa, Chiba 277-8582, Japan}

\address[utah]{Department of Physics and High Energy 
Astrophysics Institute, University of Utah,
Salt Lake City, Utah 84112, USA}

\address[eri]{Earthquake Research Institute, University of Tokyo,
Bunkyo-ku, Tokyo 113-0032, Japan}

\address[tit]{Graduate School of Science and Engineering, 
Tokyo Institute of Technology, Meguro, Tokyo 152-8551, Japan}

\address[tit2]{Interactive Research Center of Science,
Graduate School of Science and Engineering, 
Tokyo Institute of Technology, Meguro, Tokyo 152-8551, Japan}

\address[kanagawa]{Kanagawa University, Yokohama, Kanagawa 221-8624, Japan}

\address[colorado]{Department of Physics, Colorado School 
of Mines, Golden, Colorado 80401, USA}


\begin{abstract}
We performed photometric calibration of the 
PhotoMultiplier Tube (PMT) and 
readout electronics used for the new fluorescence detectors
of the Telescope Array (TA) experiment using 
Rayleigh scattered photons from a pulsed nitrogen laser beam. 
The experimental setup, measurement procedure, and results of calibration 
are described. The total systematic uncertainty of the calibration is
estimated to be 7.2\%. An additional uncertainty of 3.7\% is introduced
by the transport
of the calibrated PMTs from the 
laboratory to the TA experimental site.
\end{abstract}

\end{frontmatter}

\linenumbers

\section{Introduction}
The Telescope Array (TA) experiment is designed to observe 
extensive air showers caused by Ultra-High Energy Cosmic Rays (UHECRs),
using air fluorescence telescopes and an air shower array 
installed in the west desert of Utah, USA \cite{TA1,TA2}.  An 
important scientific objective of the TA experiment is
to measure the energy spectrum of cosmic rays
in the ultra-high energy region,
where a cutoff structure generated by the interaction of UHECRs 
with the cosmic microwave background has been predicted 
by Greissen, Zatsepin and Kuzmin (GZK) \cite{GZK1,GZK2}.  

A measurement reported by the AGASA experiment in 1998
showed a spectrum that extended beyond the expected 
GZK cutoff \cite{AGASA1,AGASA2}. 
The HiRes experiment
recently reported a strong suppression 
of cosmic ray flux \cite{HiRes} 
at around the predicted energy
of 10$^{19.7}$~eV \cite{Berezinsky}, which was
also observed by the Pierre Auger Observatory
\cite{PAO}. 

A precise measurement of the cutoff energy and the spectral 
shape around the cutoff is crucial to the identification 
of the origin of the observed structure, i.e., 
whether it is caused by the GZK effect or by some other mechanism
such as the acceleration limit of cosmic rays. 
Answering this question is an important
objective of the TA experiment.

\begin{figure}
\begin{center}
\includegraphics[width=0.7\hsize]{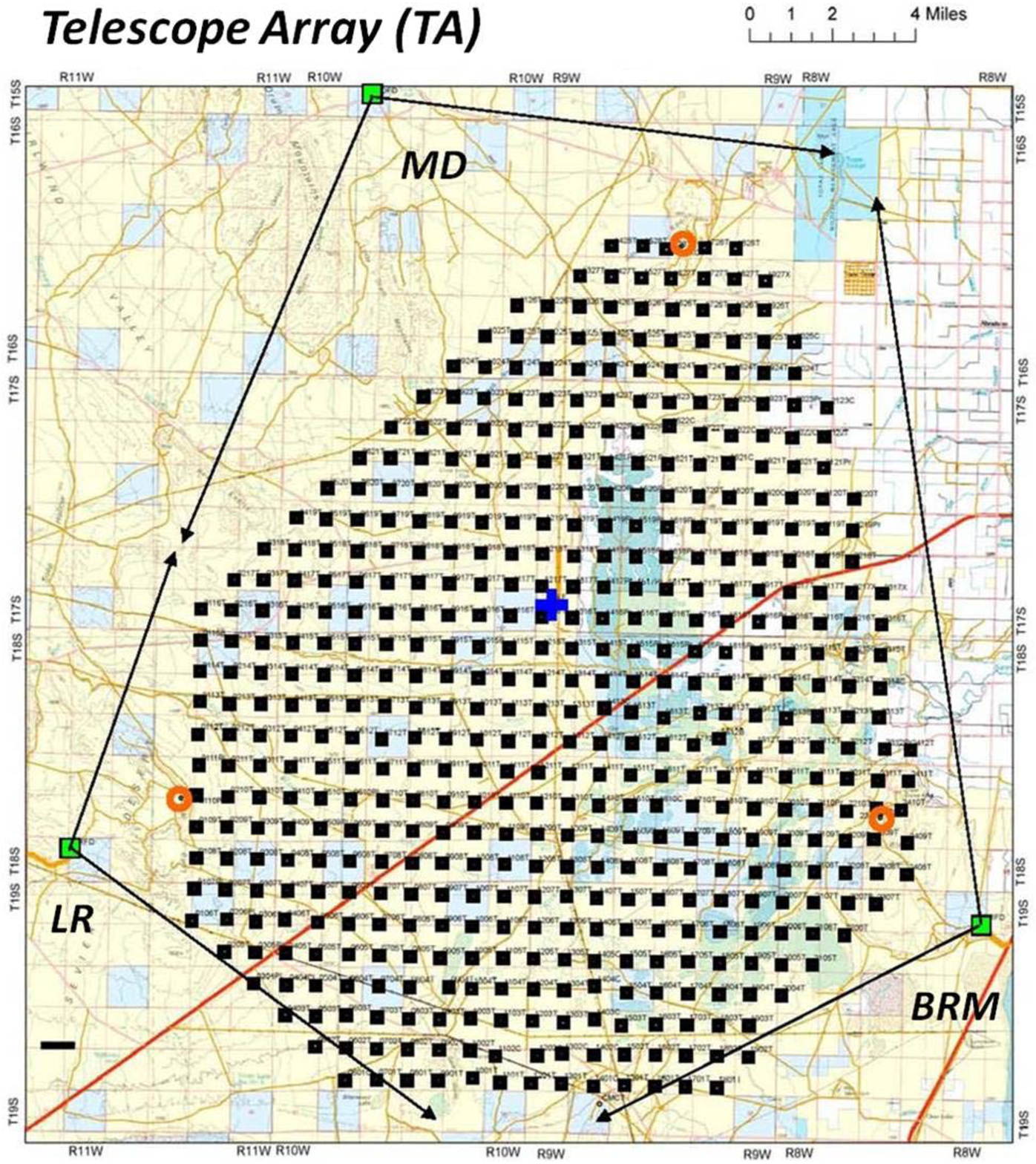}
\end{center}
\caption{Detector layout of the TA experiment. The filled squares
indicate the locations of the SDs. Three hollow squares, forming
a triangle surrounding
the SD array, show the locations of the FD telescope stations;
the extent of their azimuthal field of view is
indicated by arrows.}
\label{fig:TA_layout}
\end{figure}

The TA consists of two different types of detectors.
An air shower array covers 
a ground area of about 700~km$^2$ with 507 scintillator
Surface Detectors (SDs) 
deployed in a grid of 1.2~km spacing.
The spectral shape 
of UHECRs can be measured 
with good accuracy 
by the SD. It is fully efficient for the 
trigger and event reconstruction 
above 10$^{18.8}$~eV. 
Three Fluorescence Detector (FD) stations, each with 12-14 
fluorescence telescopes, view the sky over the surface array 
from the periphery (Figure~\ref{fig:TA_layout}).
The energies of UHECR events can be reliably determined
by the FD because it directly measures
the energy deposit in the atmosphere generated by
air showers. 

The energy determination by the FD is affected 
by several experimental uncertainties such as
the fluorescence spectrum and yield, 
the atmospheric attenuation of fluorescence photons,
the photometric calibration of the telescope, and
the missing energy carried away by high energy muons and 
neutrinos. In this paper, we address the third 
uncertainty, i.e., the photometric calibration of 
the PMTs used for the FD camera.

One of the three FD stations of the TA, Middle Drum (MD),
is located to the north of the SD array
(Figure~\ref{fig:TA_layout}).
The telescopes at the MD site are refurbished 
HiRes \cite{HiRes_calib} telescopes.
A calibration procedure similar to that employed
by HiRes using a xenon flasher was applied to the 
FDs in MD. 
The role of MD is to import the established energy 
scale of previous experiments (HiRes-1, HiRes-2, and
Fly's Eye) to the TA.

The other two FD stations, i.e., Black Rock Mesa (BRM) 
in the southeast and Long Ridge (LR) in the southwest, 
were newly produced 
for the TA experiment \cite{TA_telescope}.
A spherical mirror (diameter 3.3~m) and an 
imaging camera (16 $\times$ 16 PMT matrix) are installed
in the FDs of BRM and LR. The field of view of
one telescope is 18$^\circ$ in azimuth and 
15.5$^\circ$ in elevation.
A combination of 6 $\times$ 2 telescopes at each station provides
a field of view of 
108$^\circ$ in azimuth and 
3$^\circ$$-$33$^\circ$ in elevation.

For the new telescopes at BRM and LR, we calibrated a combination of 
PMT and readout electronics 
using a pulsed UV
light source developed specially for this purpose.
This system is composed of a pulsed nitrogen laser and a 
gas-filled chamber in which laser photons are scattered 
by the gas molecules and detected by a PMT to be calibrated. 
We call it CRAYS (Calibration using RAYleigh Scattering). 
In this paper, we describe the development of CRAYS and the 
absolute photometric calibration of the FD camera PMTs via CRAYS. 

\section{FD Camera and its Calibration}
\label{sec:camera}
\begin{figure}
\begin{center}
\includegraphics[width=0.7\hsize]{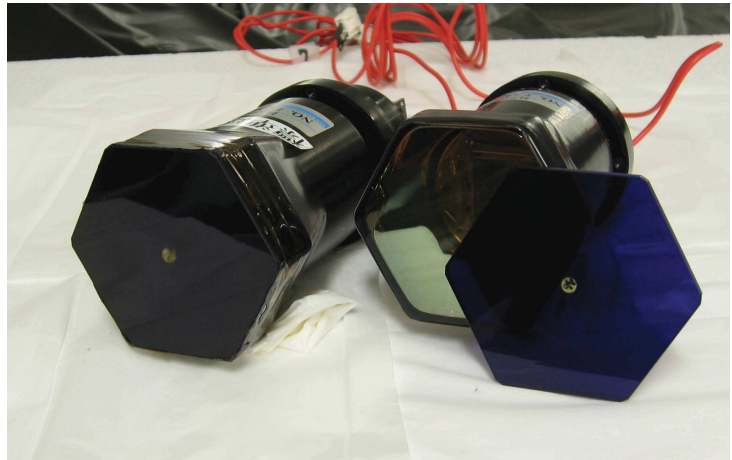}
\end{center}
\caption{PMT assembly of the TA's new FD cameras calibrated by CRAYS.
The BG3 filter contacts the PMT glass window with a thin air gap. 
On the right, the BG3 filter is removed from the PMT. An embedded YAP pulser 
can be seen at the center of the BG3 filter.}
\label{fig:PMT_photo}
\end{figure}

A photograph of the PMT assembly used for the FD camera 
is shown in Figure~\ref{fig:PMT_photo}. The PMT (R9508, Hamamatsu
Photonics) has a hexagonal photo-sensitive window
with the opposite side distance of 60~mm.
The PMT has a typical quantum efficiency of 27\%
for $\lambda$ = 337.1~nm 
(the laser wavelength) and a collection efficiency 
of 90\% as reported by the manufacturer.
The gains of all the PMTs were adjusted 
at $\sim$6.0~$\times$~10$^4$ as described later in this paper.
A UV transparent filter (BG3, Schott AG) of 4~mm thickness is 
attached to the PMT window. 
Its transmittance is measured to be 89\%
for $\lambda$ = 337~nm \cite{xenon_calib}. 

The signal from the PMT is amplified by a factor 
of 52.7 at the PMT base and is sent to 
a Signal Digitizer and Finder (SDF) module \cite{sdf} using 
a 25 m long twisted pair cable. 
The waveform is digitized by a 12-bit, 40~MHz Flash ADC
(FADC) with 2.0 V full scale. Four consecutive 
digitizations of the same input signal 
are summed together by the Field Programmable Gate Array
(FPGA) in the SDF,
and the data of 14-bit dynamic range is read out.

\begin{figure}
\begin{center}
\includegraphics[width=0.7\hsize]{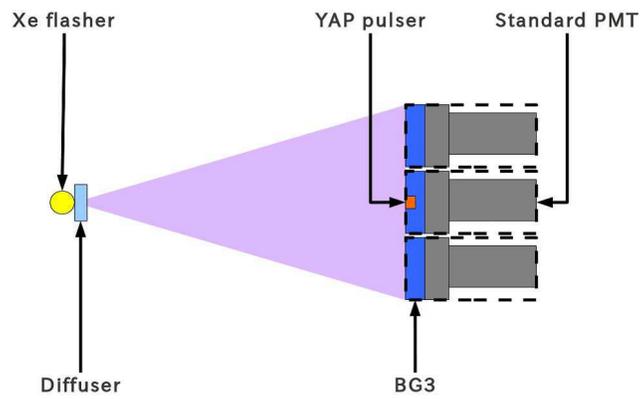}
\end{center}
\caption{Schematics of the FD camera calibration at the TA experimental site.
All the 256 PMTs in each camera 
were illuminated by the diffused xenon flasher. Only
3 PMTs were drawn in the schematics for simplicity. 
}
\label{fig:calib_schematics}
\end{figure}

The overall schematics of the FD PMT calibration 
at the TA
is shown in Figure~\ref{fig:calib_schematics}.
We calibrated 75 PMTs using CRAYS in a
laboratory at the Institute for Cosmic Ray Research (ICRR),
University of Tokyo, in Japan. 
The CRAYS-calibrated PMTs were transported to the TA experimental site in Utah, and installed into the FD cameras - one calibrated PMT at the center of each camera (Standard PMT) and another calibrated PMT toward the corner of the camera to monitor the behavior of the Standard PMT.
The same High Voltage (HV), as determined by the CRAYS
calibration at the ICRR, was applied to the Standard PMT at
the TA site. Using a diffused xenon 
flasher \cite{xenon_calib} in situ, we adjusted the HVs of all other 
PMTs (255 units) in the camera such that the gains of these PMTs 
are equal to the Standard PMT.

All the PMTs calibrated via CRAYS 
have a small YAP light pulser (diameter 4~mm) \cite{YAP}
embedded in a hole at the center 
of the BG3 filter (Figure~\ref{fig:PMT_photo}). 
The YAP is composed of a YAlO$_3$:Ce 
scintillator with 50 Bq of $^{241}$Am applied 
on the surface. The YAP generates a light flash 
(wavelength $\sim$350~nm; duration $\sim$30~ns) and produces 
approximately 450 photoelectrons in the PMT. 
The gains of the PMTs calibrated via
CRAYS in the laboratory have been monitored in the field
using the YAP signal.

\section{CRAYS}
\begin{figure}
\begin{center}
\includegraphics[width=\hsize]{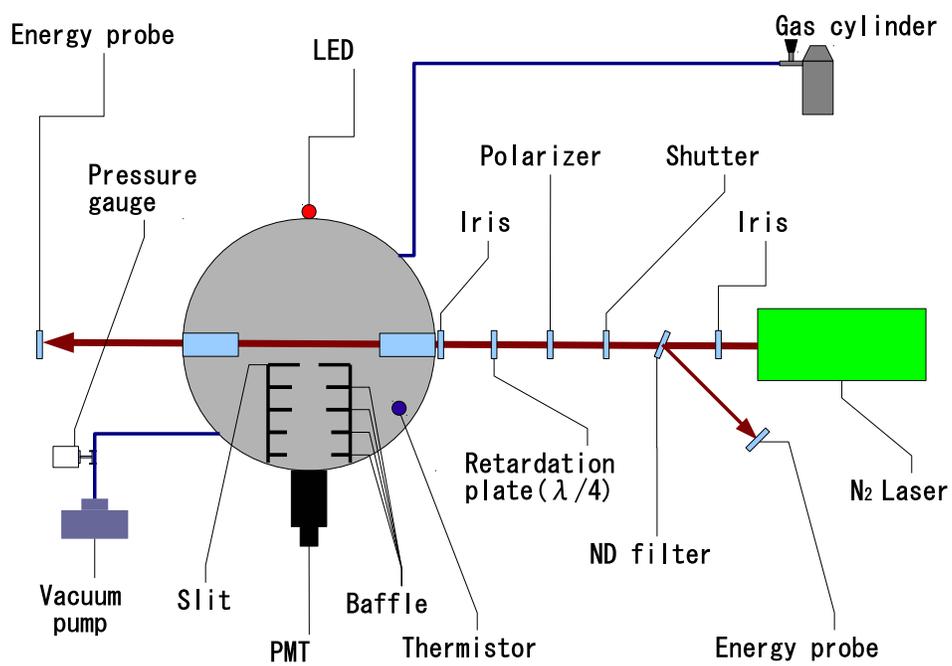}
\end{center}
\caption{Measurement setup of CRAYS.}
\label{fig:crays}
\end{figure}

The setup of CRAYS is shown in Figure~\ref{fig:crays}. 
A pulsed laser beam is directed into 
a scattering chamber filled with 
a high purity gas ($>$99.999\%) consisting of a
single molecular species, either N$_2$ or Ar.
Scattered 
photons from the beam illuminate a PMT 
viewing the chamber through a window.
Since the gas is very pure and the molecules in the gas are much 
smaller than the wavelength used, the scattering process
in the chamber is well described by molecular (Rayleigh)
scattering.
The total number of photons in the laser pulse is
calculated from the energy
measured by a calibrated energy
probe at the end of the beam line.
The number of the Rayleigh scattered photons
is calculated using the
cross-section formula,
which has been experimentally verified 
to an accuracy of $\sim$1\% \cite{Ubachs} (Sections~\ref{sec:CrossSection}).   
With a typical setup of CRAYS for nitrogen gas
(laser intensity 200~nJ; gas pressure 1000~hPa),
an intensity of approximately 80 photons/cm$^2$
is obtained on the PMT window 
(Section~\ref{PhotonAcc}).
Uncertainties of the CRAYS calibration are 0.3\% (statistical),
7.2\% (systematic), and 3.7\% (from transport to TA site)
as described in Section~\ref{sec:SysErrAll}.
We note that the same CRAYS setup was also
used with much lower laser intensity
for calibrating the IceCube PMTs 
in single photon counting mode \cite{IceCube}.

\subsection{Light Source and Optics}
We used a nitrogen laser (VSL-337ND-S, Laser Science, Inc.) 
as a light source (wavelength 337.1~nm; duration 4~ns).
The maximum energy is 300 $\mu$J 
per pulse. The wavelength
of the nitrogen laser matches that of the brightest air 
fluorescence line in the atmosphere \cite{fluo_yield}.
The diameter of the laser beam was limited to
$\sim$1~mm by a set of
irises at the exit of the laser and at the entrance of 
the scattering chamber. 
A remote-controlled shutter in the
beam line prevented the laser light from entering 
the chamber, as required.
A Neutral Density (ND) filter was used to reduce
the beam intensity. The reflected beam by the ND filter 
was measured by a pyro-electric energy probe
(Rjp-435, Laser Probe, Inc.) that monitored the relative 
intensity of the beam. 

\begin{figure}
\begin{center}
\includegraphics[width=\hsize]{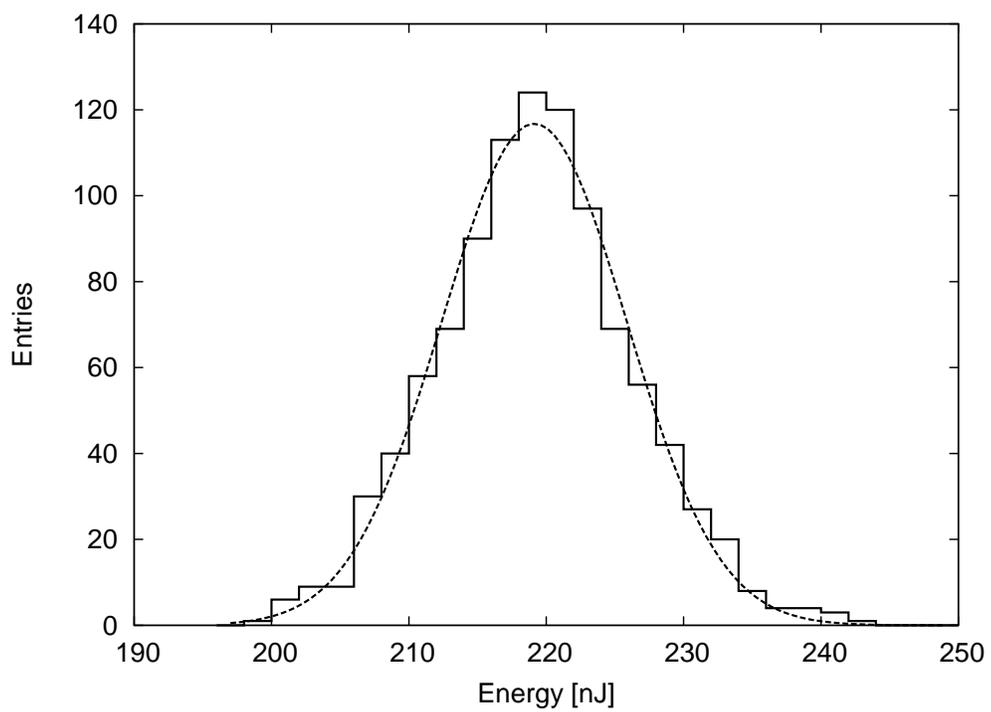}
\end{center}
\caption{Distribution of the laser energies for one calibration run.
A fit to the Gaussian is shown in the dashed line ($\sigma$/peak=0.031).}
\label{fig:laser_energy}
\end{figure}

The nitrogen laser is inherently depolarized.
To eliminate an elliptical polarization introduced 
by the ND filter, a combination of a polarizer 
and a retardation plate
($\lambda$/4) was used to convert the beam into 
a circular polarization. 
The intensity of the beam in the scattering chamber 
was measured using a silicon photodiode energy probe
(Rjp-465, Laser Probe, Inc.) placed at the end of the 
beam line. Both energy probes
were calibrated with 5\% absolute accuracy by the manufacturer.
The energy measured by Rjp-465 ranged from 190~nJ to 220~nJ
with a typical pulse-to-pulse fluctuation of 3\% as shown 
in Figure~\ref{fig:laser_energy}.

\subsection{Scattering Chamber}
The cylindrical scattering chamber has a diameter 
of 500~mm. The inner surface is anodized in black, 
and the inner wall is coated with non-reflective 
black paper to suppress stray light. The chamber was 
evacuated to $\sim$3~hPa using a membrane vacuum pump
(DAU-100, ULVAC, Inc.) before introducing the high purity
scatterer gas. The 
differential pressure of the chamber with respect to the atmospheric
pressure was monitored
by a capacitance manometer (BOC EDWARDS, Barocel 600AB) and the 
temperature inside the chamber was 
measured by a thermister thermometer.

The PMT to be calibrated was installed just outside
the chamber, as shown in Figure~\ref{fig:crays}.
The distance from the center of the chamber to
the PMT glass window was set to 312~mm. The PMT detects
photons scattered by the gas molecules near 
the center of the chamber at a scattering angle
($\theta$) of 90$^\circ$. 
The aperture of the PMT is 
limited by a slit (width 38.9~mm; height 10~mm) located
37.5~mm away from the beam line. 
The aperture is further limited by a removable mask 
installed 7~mm in front of the PMT glass window. 
Masks having a hole of 20.0~mm and 36.0~mm in diameter
exposed only the central part of 
the PMT window where the uniformity is
expected to be good.
All chamber windows are made of 
CaF$_2$ with anti-reflection coating. 
A transmittance
greater than 99\% for $\lambda$ = 337~nm
was measured by the manufacturer.

\subsection{Electronics and DAQ}
We used the same data acquisition electronics and cables
used at the TA sites as much as possible with the exception of
the high voltage power supply of the PMT.
We verified the applied HVs 
were the same at the CRAYS calibration and
at the TA sites, using a reference resistor and 
a digital multi-meter. 
Data acquisition was controlled using a PC 
that generated a trigger for the laser. The synchronization output of 
the laser was fed to the energy probes, and the energy readings 
of each laser shot were recorded by the PC.
The pressure and the temperature of the chamber were
also recorded for each calibration run. 

The waveform output from the PMT was transmitted to the 
digitizer module (SDF) installed in a VME crate. 
The synchronization
signal from the laser was recorded by the SDF to define 
the signal integration interval in the off-line analysis.
For YAP data recording, a trigger was generated in the SDF
by the YAP signal itself. The DAQ rate 
was approximately 20 Hz for the CRAYS run and 50 Hz for
the YAP run.

\section{Performance Check}
\label{sec:performance}
\begin{figure}
\begin{center}
\includegraphics[width=\hsize]{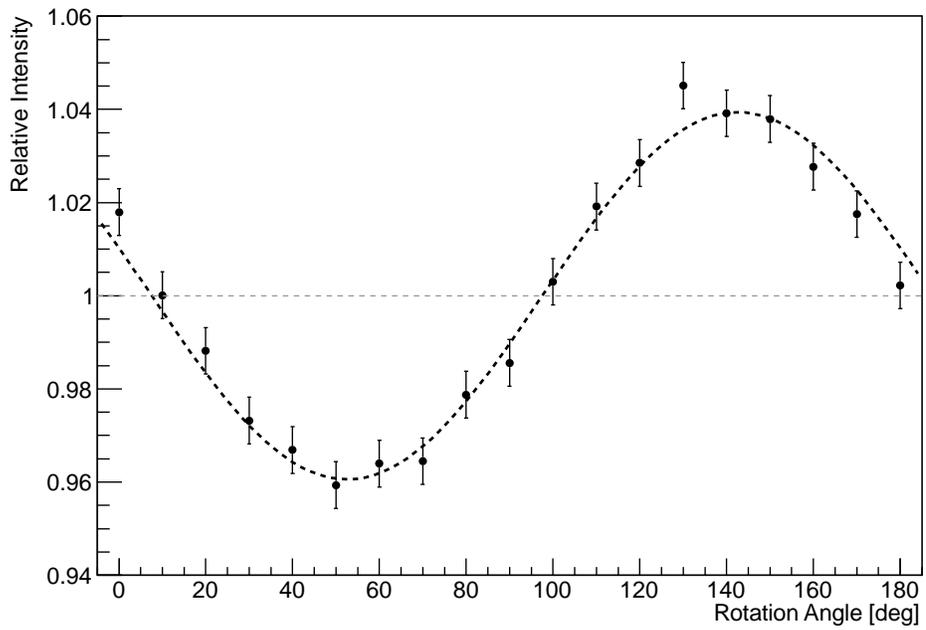}
\end{center}
\caption{Relative change of the measured laser energy with 
respect to the polarizer rotation angle. 
A fit to the sinusoidal function is shown in the dashed line.}
\label{fig:polarization}
\end{figure}

Before using CRAYS for calibration, we made the 
following investigations
to ensure that the photons detected 
by the PMT originated from the Rayleigh scattering of 
the laser beam and that the background photon was
under control.
First, the polarization of the beam was measured 
by temporarily inserting a rotatable polarization plate and 
recording the output of the energy probes 
at different rotation angles. In Figure~\ref{fig:polarization},
the relative intensity of the laser beam measured by
the downstream energy probe
is plotted with respect to
the change of the polarizer rotation angle $\phi$.
A fit to the sinusoidal curve
\begin{eqnarray}
\label{eq:PolPhi}
1 + A \sin 2 (\phi + \phi_0)
\end{eqnarray}
was made with an amplitude $A$ and a phase $\phi_0$ as free parameters. 
The obtained values, $A$ = $-$0.04 and $\phi_0$ = $-$8$^\circ$, indicate
an elliptical polarization of 4\%
in the axis 37$^\circ$ away from the
vertical-upward direction.
An effect of the polarization on the number of expected Rayleigh scattered
photons in the CRAYS setup is described in Section~\ref{sec:Uncertainty}.

\begin{figure}
\begin{center}
\includegraphics[width=\hsize]{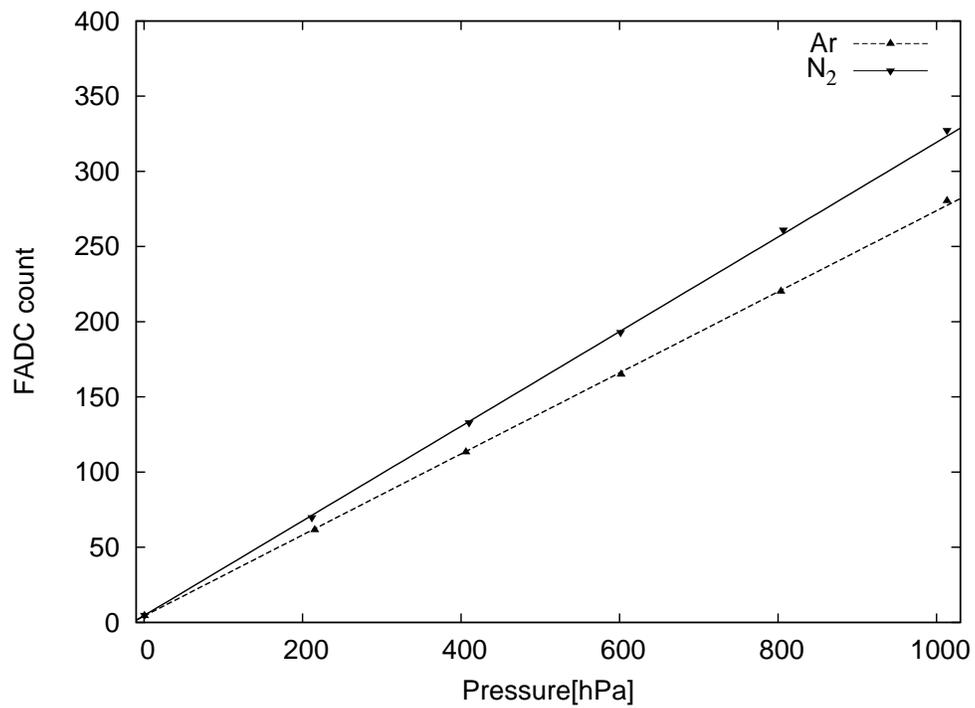}
\end{center}
\caption{Integrated FADC counts generated by the photons scattered
from the laser beam with respect to the change in gas pressure. 
A linear fit is shown
in the solid line (nitrogen) and in the dashed line (argon).
}
\label{fig:pressureVScount}
\end{figure}

Next, the amount of the scattered photons and the 
PMT responses were measured by changing 
the pressure of the gas between 3 and 1013~hPa. 
The integration of the FADC signal 
and the pedestal subtraction were done in the same manner
as described in Sections~\ref{Integration} and~\ref{Results}.  
The result of the measurements for nitrogen and argon gas are shown in 
Figure~$\ref{fig:pressureVScount}$. Good linearities of the PMT
output with respect to the change of the gas pressure
were obtained both for nitrogen and argon. 
The argon to nitrogen ratio (Ar/N$_2$) was
0.857 $\pm$ 0.007 from a linear fit to the measured FADC
counts and taking a ratio of the two slopes.
The measured ratio is in a good agreement with
the theoretical cross-section
calculation, which predicts a value of 
0.849 (Section~\ref{sec:CrossSection}).

A signal of 16-photons-equivalent was detected
in the vacuum setup. This is about 1.9\% of the Rayleigh 
scattered photons for the laser energy
of 200~nJ, measured with the PMT mask of 36~mm$\phi$
(nitrogen gas; pressure, 1000~hPa). 
This background without
scatterer molecules in the CRAYS chamber
was attributed to the stray light
generated by reflection of the laser 
by beam line elements such as the CaF$_2$ window and 
the energy probe. The background amount
was stable during the calibration runs, and its contribution
to the PMT signal
was subtracted in the data analysis.

\begin{figure}
\begin{center}
\includegraphics[width=\hsize]{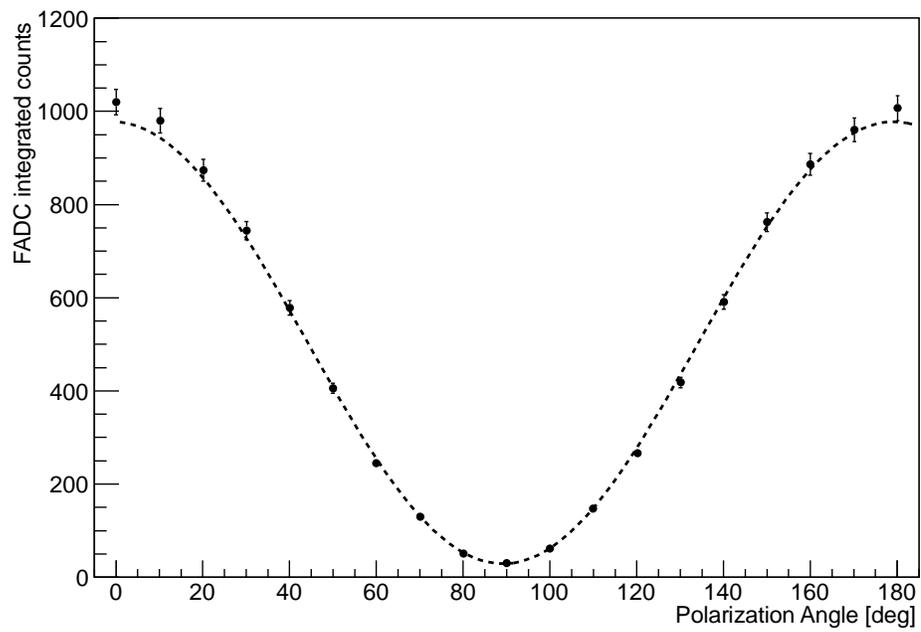}
\end{center}
\caption{Integrated FADC counts generated by the photons scattered
from the polarized laser beam with respect to the change of polarization
angle. A fit to the sinusoidal function is shown in the dashed line.}
\label{fig:RayleighPhiDep}
\end{figure}

Finally, a linear polarization was artificially introduced in the beam 
line using the rotatable polarization plate, and the PMT signal was
measured for different polarization angles. The measurement was made
for nitrogen gas. 
Figure~\ref{fig:RayleighPhiDep} shows a change of the
integrated FADC count for different settings of the
rotation angle ($\phi$) of the polarization plate
between 0$^\circ$ and 180$^\circ$, where
$\phi$ is defined to be zero in the vertical-upward direction.
The data points are
fitted with a sinusoidal function \cite{Miles},
\begin{eqnarray}
\label{eq:LinPolPhi}
A \left[ \frac{1 + \rho_0}{2 + \rho_0}  - \frac{1 -
\rho_0}{2 + \rho_0} \cos 2 (\phi + \phi_0) \right] + B
\end{eqnarray}
where an amplitude $A$, a background $B$, and a phase
$\phi_0$ are free parameters, and a depolarization ratio, $\rho_0$,
is introduced as a constant of 0.022 (Section~\ref{sec:SysErrAll}).
We obtained
$A$ = 980.1, $B$ = 8.1, $\phi_0$ = $-$89.2$^\circ$ with
$\chi^2$/NDF = 22.9/16. 
The minimum value at $\phi = -\phi_0$
is 3.0 \% of the maximum value, which is attributed
to a depolarization effect of diatomic 
nitrogen gas (2.2 \%) and the unpolarized background (0.8\%).

\section{Calibration Procedure}
We calibrated a total of 75 PMT assemblies with CRAYS.
The procedure is listed below.
\begin{enumerate}
\item
A relation between the PMT gain and the applied HV
was measured by pulsing a UV LED, installed in the chamber opposite
to the PMT (Figure~\ref{fig:crays}). 
A set of LED runs were carried out in a range
between $-$700 V and $-$1250 V. 
The integrated FADC counts X and the HV setting Y are well
fitted with a function X = $\alpha$Y$^\beta$, yielding 
a measurement of the parameter $\beta$~=~8.1$~\pm$~0.4(rms).

\item
Next, several laser runs were made for each
PMT to determine the HV setting for
the calibration. The scattering chamber was filled with 
nitrogen gas ($\sim$1010~hPa) and a PMT 
mask (36~mm$\phi$) was attached. 
The HV to be applied to each PMT was tuned 
iteratively using the gain-HV relation (step-1)
such that all the calibrated PMTs had approximately
the same integrated FADC counts
($\sim$360 counts for a 200~nJ laser pulse).
The average of the resultant HV settings 
for the 75 PMTs was $-$870 $\pm$ 50(rms) V.

\item
By applying the HV determined (step-2),
three CRAYS laser runs 
were carried out to measure
the PMT response 
with three different PMT mask conditions: 
20~mm$\phi$, 36~mm$\phi$, and no mask. 
\item
After the laser calibration, the YAP data was recorded
with the same HV setting for future reference.
\end{enumerate}

For each CRAYS run, we collected the data of 2000 laser shots: 
1000 shots with shutter-open and 1000 shots with shutter-closed. 
We alternated the shutter status every 100 laser shots. 
The shutter-closed data was used
to subtract the electrical noise synchronized with the 
laser shots. The energy probe readings were recorded for 
each laser shot. The temperature and pressure of the gas inside
the chamber were continuously monitored. The YAP data was also taken
for 2000 events. 

The temperature in the laboratory where the CRAYS setup was installed 
was maintained at 25 $\pm$ 1$^\circ$C during the measurement, and 
the absolute atmospheric pressure was measured
by a mercury pressure gauge for each calibration run. 

\section{Data Analysis}
\subsection{Photon Acceptance}
\label{PhotonAcc}
The cross-section of Rayleigh scattering in 
nitrogen gas at $\lambda$ = 337.1~nm is given
by the expression (Section~\ref{sec:CrossSection})
\begin{eqnarray}
\label{eq:dsdth}
\frac{d\sigma_R}{d\Omega}=\frac{3}{16\pi}
(1+\cos^2\theta)~\times~3.50~\times~10^{-26}~{\rm cm}^2
\end{eqnarray}
The molecular density $N$ of the scatterers 
can be determined from the equation of state for the ideal gas,
\begin{eqnarray}
PV = NRT
\end{eqnarray}
where $P$ is the pressure,
$V$ is the volume, 
$T$ is the temperature [K], and 
$R$ is the gas constant having a value of 8.31 [J/K/mol]. 
For nitrogen gas at 1000~hPa and 25$^\circ$C, 
$N$ = 2.43 $\times$ 10$^{19}$~cm$^{-3}$. The minor correction 
for Van der Waals gas can be neglected for our purpose.

A pulse of 200~nJ nitrogen laser beam includes 
3.39 $\times$ 10$^{11}$ photons. With a Rayleigh 
scattering cross-section of 
3.50 $\times$ 10$^{-26}$~cm$^2$, the number of Rayleigh scattered 
photons along the beam line inside the chamber 
is 1.43 $\times$ 10$^7$.  

We performed ray tracing of Rayleigh scattered photons 
in the chamber in order to estimate the number of photons accepted
by the PMT. The Rayleigh scattered photons were produced along 
the beam line with a scattering angle dependence of 
1 + cos$^2$$\theta$ and with uniform azimuthal angle dependence.
The generated photons were allowed to enter the PMT directly
or with one scattering on a chamber element such as the 
inner wall or the baffles. The shadow of the YAP embedded in 
the BG3 filter was also taken into account. 

The ray tracing MC simulation showed that
the average number of photons that reached 
the PMT window was 
823 for nitrogen gas at 1000~hPa
with a PMT mask of 36~mm$\phi$, and 
the laser intensity of 200~nJ.
An effective length of 48.8~mm of the laser beam line 
near the chamber center was seen from the PMT. The photons entered normal 
to the PMT window within 8$^\circ$,
making a nearly uniformly 
irradiated circular area (diameter 36.6~mm) on the PMT window.
  
The effect of stray light originating from the Rayleigh scattering
by the beam line was estimated by changing reflection coefficient
of the chamber inner walls. We used a measured
reflectivity of 0.023 for the chamber inner wall. For this value and 
assuming mirror scattering,
three photons on average were detected after
a single scattering on the chamber wall
in addition to the 823 photons of direct incidence.
The number was less than one
when a random (isotropic) scattering was assumed. 
Because the scattering is expected to be close to
Lambertian on the black paper
covering a major part of the chamber wall, we concluded 
that the effect of stray light originating from the Rayleigh 
scattering in the beam line is negligible. The effect of
multiple scattering on the chamber wall was also
tested to be negligible.

\subsection{Waveform Integration}
\label{Integration}
\begin{figure}
\begin{center}
\includegraphics[width=\hsize]{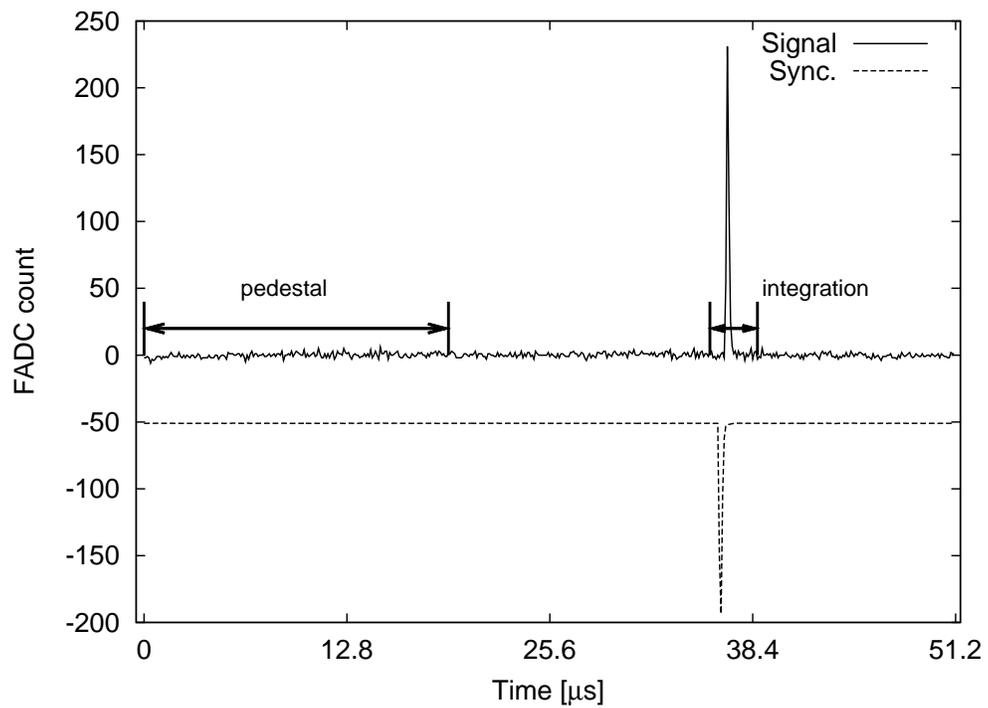}
\end{center}
\caption{Typical PMT waveform from CRAYS.
The time intervals for the pedestal 
determination and the signal integration are indicated.
The laser synchronization signal (dashed line) 
is inverted.}
\label{fig:waveformCRAYS}
\end{figure}

A typical digitized PMT waveform is shown 
in Figure~\ref{fig:waveformCRAYS}. 
A time interval of 51.2 $\mu$s was recorded centered on
the PMT signal. 
The PMT signal was detected within 100~ns of 
the laser synchronization signal (Figure~\ref{fig:waveformCRAYS}).
We determined the range of signal integration to be 
1 $\mu$s before and 2 $\mu$s 
after the peak of the synchronization signal. The pedestal level
was evaluated as an average of 19.2 $\mu$s duration at the beginning 
of the recorded waveform, and it was subtracted before integration.
The accidental overlap of the YAP signal in the pedestal 
evaluation interval was low ($\sim$0.1\%), but when it happened,
it was recognized by looking at the pedestal histogram, 
and removed. 

\begin{figure}
\begin{center}
\includegraphics[width=\hsize]{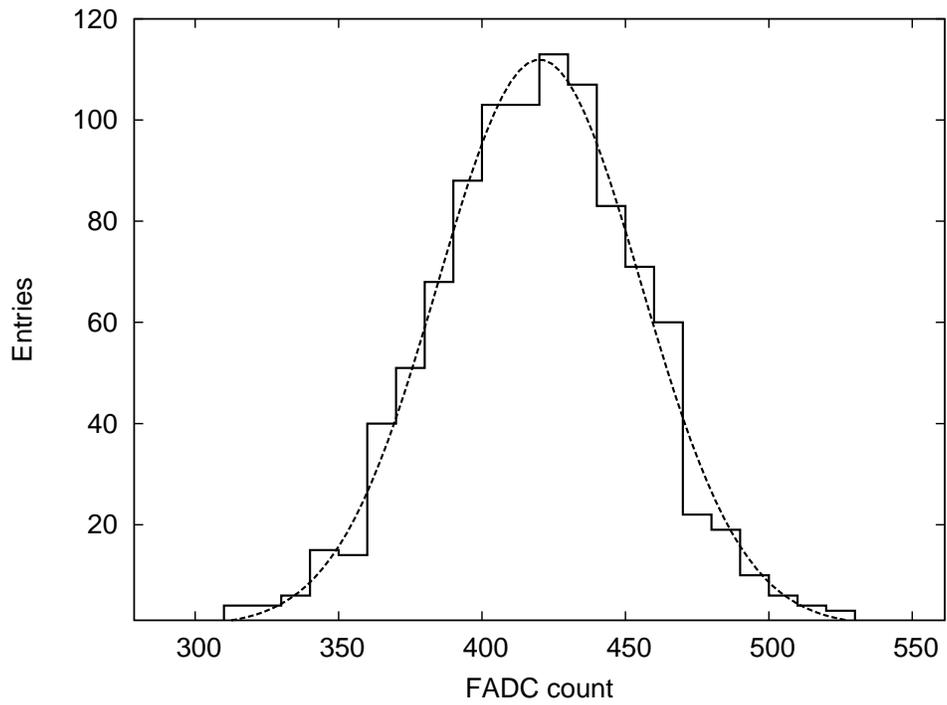}
\end{center}
\caption{Distribution of the $\Sigma_{\rm ADC}$
for a CRAYS calibration run.
A fit to the Gaussian
is shown in the dashed line ($\sigma$/peak=0.085).
}
\label{fig:integrated_FADC_count}
\end{figure}

A typical distribution of integrated PMT signals is shown
in Figure~\ref{fig:integrated_FADC_count}, after correcting
the FADC signal for the
shot-to-shot fluctuation in the laser energy (normalized to
the average energy). 

The signal resolution defined by $\sigma$/peak of the distribution
is 8.5\%, which is attributed to the statistical fluctuation of 
photoelectrons received by the first dynode ($\sim$7.0\%), the single
photoelectron resolution ($\sim$3\%), and the electronics noise
contribution ($\sim$4\%).

\section{Results}
\label{Results}
The photometric calibration constant $C$ of the PMT-electronics system
is defined as $C = N_\gamma/\Sigma_{\rm ADC}$ where
$N_\gamma$ means the total 
number of photons striking the PMT 
sensitive area and $\Sigma_{\rm ADC}$ means 
the sum of the recorded FADC 
counts. We used the measured laser energy, gas temperature, and 
pressure for calculating
the N$_\gamma$ to be detected by the PMT. 
We subtracted the contribution of the 
shutter-closed state from the shutter-open state as a background when calculating
$\Sigma_{\rm ADC}$.

The following set of parameters were obtained for each calibrated PMT.
\begin{enumerate}
\item operation HV setting
\item calibration constant, $C$, with 36~mm$\phi$ PMT mask
\item $\Sigma_{\rm ADC}$ with 20~mm$\phi$ PMT mask
and without PMT mask, normalized to 200~nJ laser energy. 
\item $\Sigma_{\rm ADC}$ of the YAP pulser
\end{enumerate}  
 
\begin{figure}
\begin{center}
\includegraphics[width=\hsize]{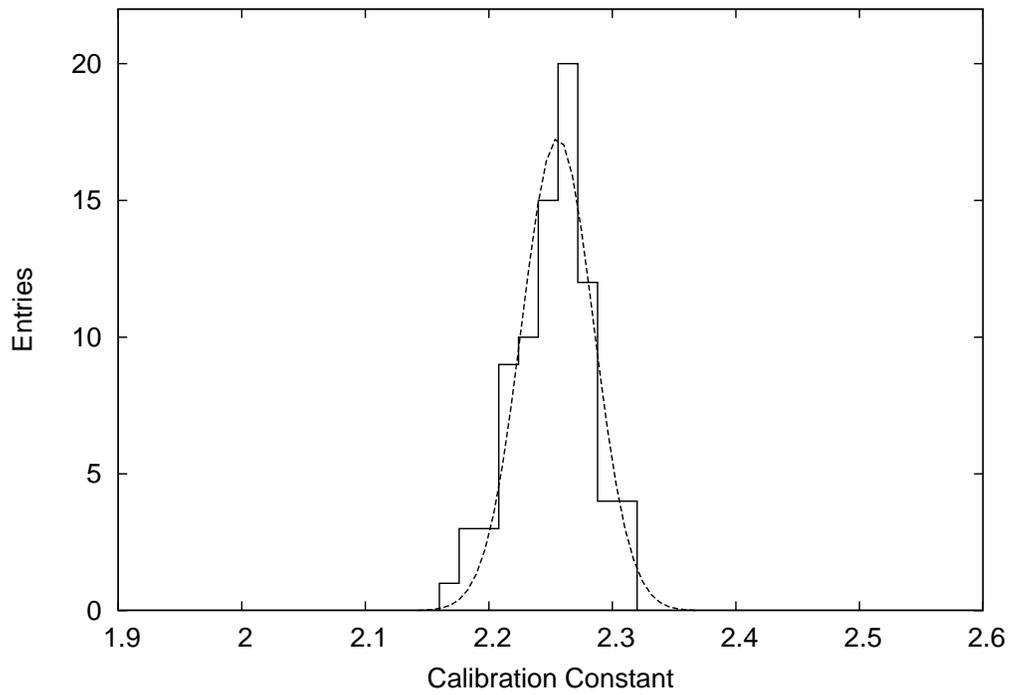}
\end{center}
\caption{Distribution of the calibration constant, $C$,
for 75 calibrated PMTs (36~mm$\phi$ mask).
A fit to the Gaussian
is shown in the dashed line (peak=2.256, $\sigma$=0.0291).
}
\label{fig:calib_36ph_20ph}
\end{figure}

The distribution of $C$ for all the 75 calibrated PMTs 
with 36~mm$\phi$ PMT mask is shown
in Figure~\ref{fig:calib_36ph_20ph}.
The statistical accuracy of the calibration is
better than 0.3\%. 
These values
are used in the air shower analysis of the TA 
as calibration constants.
The average of 2.25 [photons/FADC count] 
in  Figure~\ref{fig:calib_36ph_20ph} corresponds to
the PMT amplification of 6.0 $\times$ 10$^4$ using all the known optical
and electrical parameters of the PMT camera system
(Section\ref{sec:camera}).

\begin{figure}
\begin{center}
\includegraphics[width=\hsize]{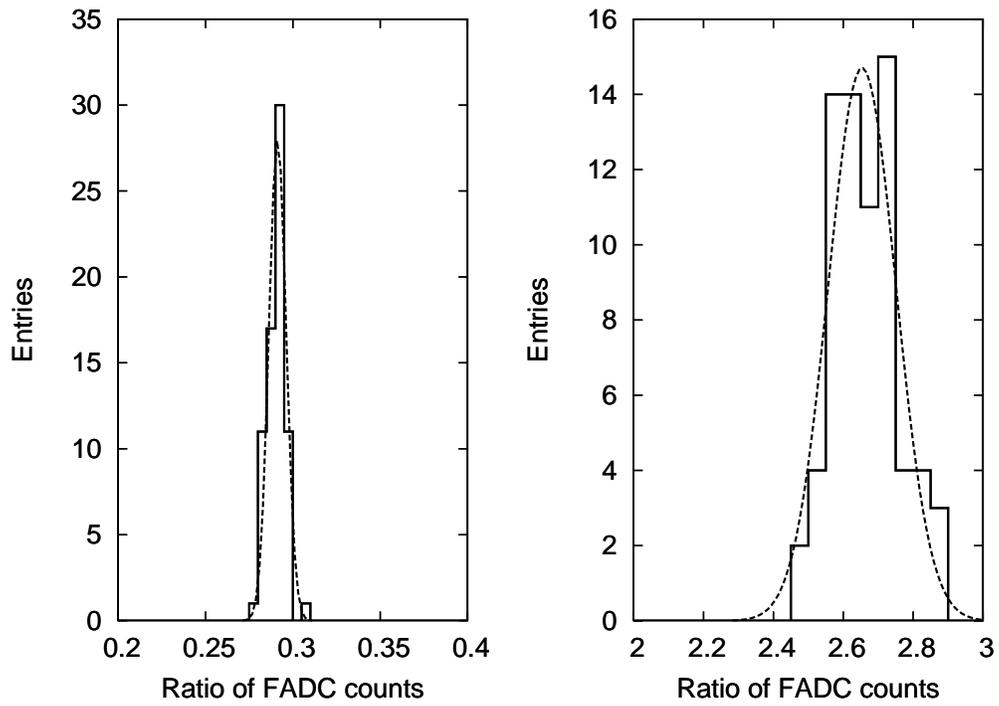}
\end{center}
\caption{Distribution of the ratio of
$\Sigma_{\rm ADC}$ for 75 PMTs;
20~mm$\phi$-mask/36~mm$\phi$-mask (left) and
no-mask/36~mm$\phi$-mask (right). 
A fit to the Gaussian
is shown in the dashed line 
(peak = 0.291, $\sigma$=0.050 for the left, and
peak = 2.65, $\sigma$=0.097 for the right).
}  
\label{fig:calib_masks}
\end{figure}

The ratios of $\Sigma_{\rm ADC}$
obtained for different mask settings
are shown in Figure~\ref{fig:calib_masks} for 75 PMTs together
with the Gaussian fitting.
The expected values of these ratios are
0.294 (20~mm$\phi$-mask/36~mm$\phi$-mask) and
2.73 (no-mask/36~mm$\phi$-mask)
from the 2-dimensional sensitivity 
scanning of the PMT window \cite{xenon_calib}.
The fitted peaks of Figure~\ref{fig:calib_masks}
are 0.291 and 2.65 respectively, and the measurements
agreed with the expectation within 3\%. The widths
($\sigma$/peak) of the two distributions, 3.4\% for
no-mask/36~mm$\phi$-mask and 1.7\% for
20~mm$\phi$-mask/36~mm$\phi$-mask,
indicate the level of uniformity of the 
photo-sensitive area among the calibrated PMTs.
The accuracy of the no-mask/36~mm$\phi$-mask ratio is 
relevant for transmitting the calibration 
of the Standard PMT to other PMTs in a given camera,
which were used for the observation without any mask,
by using a diffused 
xenon flasher in situ.

\section{Systematic Uncertainties}
\label{sec:SysErrAll}
\subsection{Rayleigh Scattering Cross-Section}
\label{sec:CrossSection}
The total Rayleigh scattering cross-section $\sigma_R$
for a single molecule is given by (e.g. \cite{Bucholtz})
\begin{eqnarray}
\sigma_R(\nu) = \frac{24 \pi^3 \nu^4}{N^2} \left( \frac{ n_{\nu}^2 - 1 }{n_{\nu}^2 + 2} \right)^2F_K(\nu) \label{eq:RCrossSection}
\end{eqnarray}
where $\nu$ is the wavenumber [1/wavelength], $N$ is the
molecular density, $n_{\nu}$ is the refractive index, and
$F_K(\nu)$ is the King correction factor accounting for
the anisotropy of scatterings by non-spherical molecules. 
In order to use the equation~(\ref{eq:RCrossSection}), the values of
$n_{\nu}$ and $N$ should be chosen in a consistent way 
(i.e. values under a same condition in temperature and pressure)
because of the relation $(n_{\nu}^2 - 1)/(n_{\nu}^2 + 2) \propto N$ \cite{Jackson}. We use 
$n_{\nu}$ values at NTP (normal temperature and pressure,
$T = 273.15~{\rm K}$ and $P = 1013.25~{\rm hPa}$), and
we take $N = 2.69 \times 10^{19} \,~{\rm cm^{-3}}$ \cite{CODATA}.

\begin{figure}
\begin{center}
\includegraphics[scale=0.7]{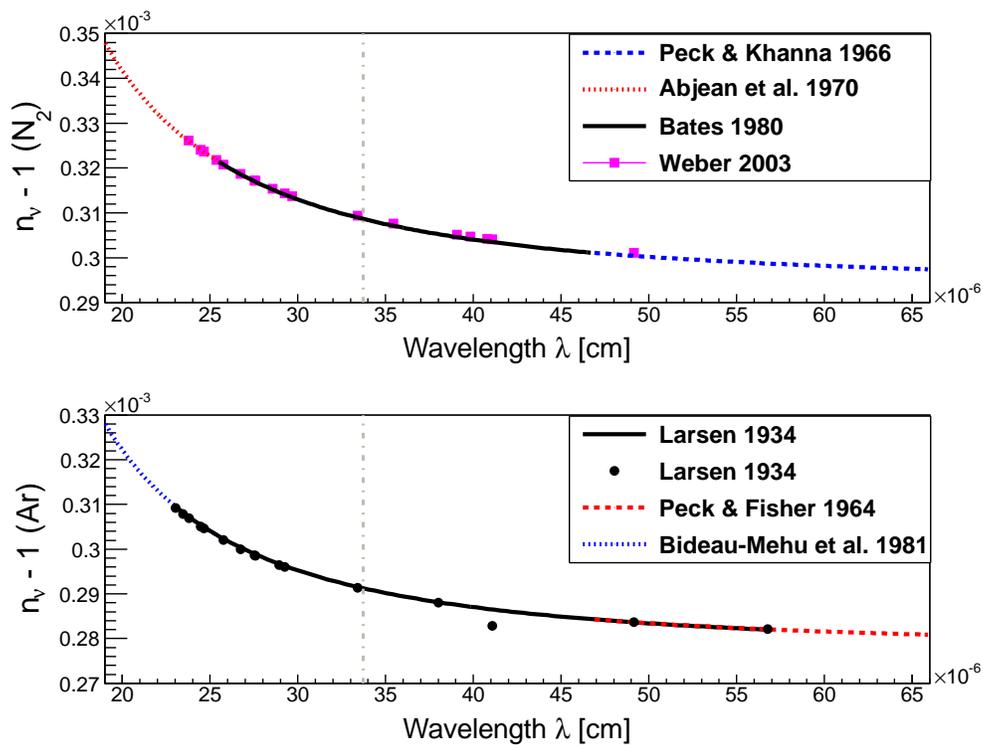}
\end{center}
\caption{Refractive indices of nitrogen (upper) and argon (lower).}
\label{fig:refindex}
\end{figure}

Peck and Khanna \cite{PeckKhanna} gave an empirical formula
for the refractive index of nitrogen at NTP
in the wavelength range $468-2060$~${\rm nm}$ as
\begin{eqnarray}
10^8 (n_{\nu} - 1) = 6855.200 + \frac{3.243157 \times 10^{14}}{1.44 \times 10^{10} - \nu^2} \label{eq:Peck}
\end{eqnarray}
where $\nu$ is in [1/cm].
Abjean {\it et al.} \cite{Abjean} made a similar expression for a shorter wavelength range $181-254$~${\rm nm}$,
\begin{eqnarray}
10^8 (n_{\nu} - 1) = 6998.749 + \frac{3.233582 \times 10^{14}}{1.44 \times 10^{10} - \nu^2} \label{eq:Abjean}
\end{eqnarray}
Bates \cite{Bates} gave an interpolation to cover the intermediate range for $254-468$~${\rm nm}$ in the same form as (\ref{eq:Peck})
and (\ref{eq:Abjean}) as
\begin{eqnarray}
10^8 (n_{\nu} - 1) = 5989.242 + \frac{3.3632663 \times 10^{14}}{1.44 \times 10^{10} - \nu^2} \label{eq:Bates}
\end{eqnarray}
This well reproduces the data in the literature \cite{CRC} in
$238-490$~${\rm nm}$.  These formulae and data are shown
in Figure~\ref{fig:refindex}.

Larsen \cite{Larsen,Leonard} measured the refractive index of argon 
at NTP in $230 -567$~${\rm nm}$ and gave an expression
\begin{eqnarray}
\frac{3}{2} \left( \frac{n_{\nu}^2 - 1}{n_{\nu}^2 + 2} \right) & = &1.2098 \times 10^6 \left( \frac{0.208972}{0.87882 \times 10^{10} - \nu^2} \right. \nonumber \\
& {} & {} + \left. \frac{0.208972}{0.9100 \times 10^{10} - \nu^2} + \frac{4.925837}{2.69636 \times 10^{10} - \nu^2} \right) \label{eq:Larsen}
\end{eqnarray}
where $\nu$ is in [1/cm]. This is also shown in Figure~\ref{fig:refindex},
together with the measurements in different wavelength ranges given
in \cite{PeckFisher} and \cite{BM}.
				
The empirical formulae for $n_{\nu}$ of nitrogen and argon well fit 
the data in the wide range, including the wavelength of our interest
$\lambda = 337.1$~nm. For our calculation, we use the equation~(\ref{eq:Bates})
for nitrogen and (\ref{eq:Larsen}) for argon, which are evaluated
as $n_{\nu}({\rm N}_2) - 1 = 3.0865 \times 10^{-4}$ and $n_{\nu}({\rm Ar}) - 1 = 2.9119 \times 10^{-4}$, respectively.

The values of the King correction factor for nitrogen have been derived from the measurements by Bridge and Buckingham \cite{Bridge}, Alms {\it et al.}
\cite{Alms}.
A widely used dispersion relation  for the King correction factor of nitrogen was given by Bates \cite{Bates} using these data and the calculations by Oddershede and Svendsen \cite{Oddershede},
\begin{eqnarray}
F_K(\nu) = 1.034 + 3.17 \times 10^{-12} \nu^2 \label{eq:BatesKingC}
\end{eqnarray}
where $\nu$ is in [1/cm]. Since argon is of monoatomic
molecule, $F_K({\rm Ar}) = 1$ is expected. The measurement by Rudder and Bach  \cite{Rudder} showed that
the degree of depolarization is $\sim 10^{-5}$, and the deviation of $F_K({\rm Ar})$ from unity is $3 \times 10^{-5}$ \cite{Frohlich,Srivastava}.

Using the $n_{\nu}$ and $F_K(\nu)$ values described above,
we obtained the total Rayleigh scattering cross-sections of
nitrogen and argon at $\lambda = 337.1$~nm under NTP, as
\begin{eqnarray}
\sigma_R({\rm N}_2) = 3.50 \times 10^{-26} \, {\rm cm^2}, \;\; \sigma_R({\rm Ar}) = 3.00 \times 10^{-26} \, {\rm cm^2} \label{eq:CScalculated}
\end{eqnarray}
We used these cross-sections in our ray-tracing simulation
of scattered laser photons in the CRAYS chamber (equation~(\ref{eq:dsdth})).
The accuracies of 
$\sigma_R({\rm N}_2)$ and $\sigma_R({\rm Ar})$ which
come from uncertainties in $n_{\nu}$ and $F_K$ (for nitrogen) are 
$1\%$ and $0.3\%$, respectively (see also \cite{Ubachs}).

The argon to nitrogen ratio is 
$\sigma_R({\rm Ar})/\sigma_R({\rm N}_2)  = 0.858$.
The ratio that CRAYS measures at $\theta = 90^\circ$
becomes 0.849, being slightly affected by the modification of the
differential cross section for diatomic molecules
(N$_2$) \cite{Miles}.

\begin{figure}
\begin{center}
\includegraphics[scale=0.65]{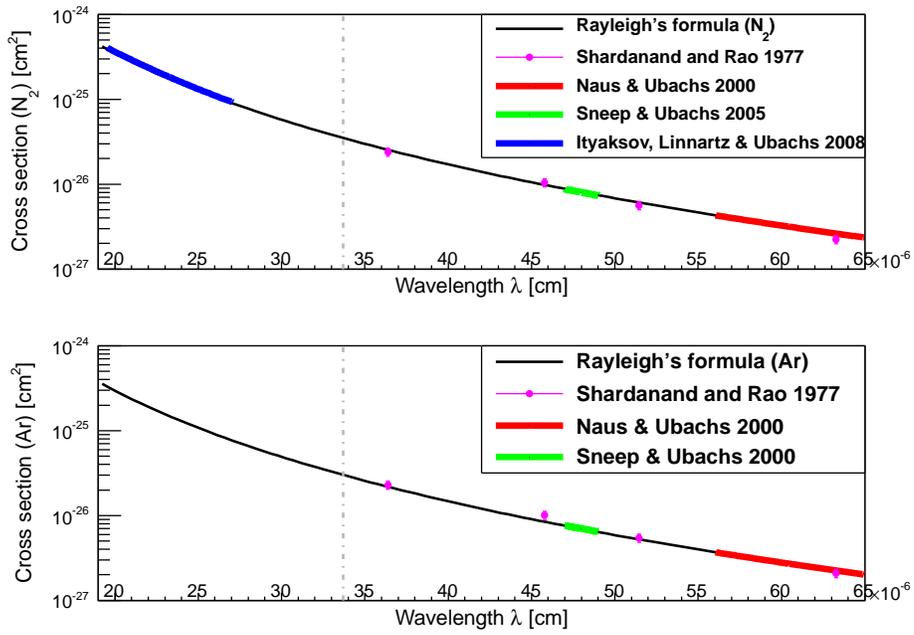}
\end{center}
\caption{Rayleigh scattering cross-sections. The solid line is
calculation by the formula~(\ref{eq:RCrossSection}) using the refractive
indices and the King correction factor given in \cite{Bates}.
Fits for the
experimental data given in the
literature \cite{Shardanand,Ubachs,NausUbachs2,Sneep,Ityaksov}
are shown in different colors. 
}
\label{fig:CrossSectionM}
\end{figure}

Experimental verifications of the Rayleigh cross-section
formula~(\ref{eq:RCrossSection}) for gases in optical and ultraviolet wavelengths
is rather scarce. An old measurement by Shardanand and Rao
\cite{Shardanand} gave cross-section values for nitrogen and argon 
at 5 wavelengths from $363.8$~nm to $632.8$~nm, which are in agreement
from expectations within $1 \sim 5\%$ (Figure~\ref{fig:CrossSectionM}).
Naus and Ubachs firstly carried out a modern laboratory laser measurement
of Rayleigh scattering cross-sections of nitrogen and argon in $560-650$~nm
with the cavity-ringdown technique \cite{Ubachs,NausUbachs2}. They
compared their measured values of cross-section and the expectations from
the formula~(\ref{eq:RCrossSection}) with $n_{\nu}$ evaluated with
(\ref{eq:Peck}) (\ref{eq:Larsen}) and $F_K(\nu)$ by (\ref{eq:BatesKingC}), and
concluded that the measured and the calculated cross-sections agree
within an experimental uncertainty of $1\%$. They also gave an empirical
expression for the Rayleigh cross-section in a form
\begin{eqnarray}
\sigma_R(\nu) = \bar{\sigma} \nu^{4 + \epsilon} \label{eq:CSempirical}
\end{eqnarray}
By fitting their measured values to (\ref{eq:CSempirical}) they obtained 
$\bar{\sigma} = 22.94 \times 10^{-45}$ and $\epsilon = 62.4 \times 10^{-3}$ for nitrogen, and 
$\bar{\sigma} = 19.89 \times 10^{-45}$ and $\epsilon = 61.5 \times 10^{-3}$ for argon \cite{Ubachs}. This experiment was followed by
the measurements in shorter wavelengths, as Sneep and Ubachs
in $470-490$~nm \cite{Sneep}, and Ityaksov, Linnartz and Ubachs
in $198-270$~nm  \cite{Ityaksov}.
Although there are few cross-section data available in the very
vicinity of our interest, $\lambda = 337.1$~nm,
the measured values both in the shorter and the longer wavelength
ranges are in excellent agreement with the expectation
from (\ref{eq:RCrossSection})
within $\sim$ 1\%, and there is no evidence of non-validity of
(\ref{eq:RCrossSection}) at 337.1~nm.

\subsection{Uncertainty of CRAYS}
\label{sec:Uncertainty}
A list of systematic uncertainties for the calibration
constant, $C$ (36~mm$\phi$), obtained by CRAYS is given
in Table~\ref{tb:sys_error}.
The calibration of PMTs with CRAYS
is fully dependent on an evaluation of the
total and differential cross-sections of
Rayleigh scattering, $\sigma_R$ and $d\sigma_R/d\Omega$,
and its modification due to the polarization of the incident
laser beam. As described in Section~\ref{sec:CrossSection},
the direct measurement of $\sigma_R$ agrees
with the calculation within $\sim$1\%
in the shorter and in the longer wavelength
ranges around 337.1~nm.
Using CRAYS,
 we
 measured the argon-to-nitrogen ratio
at $\lambda$ = 337.1~nm and showed that the calculation and
the measurement of the ratio agree also within 
1\% (Section~\ref{sec:performance}).
This measurement gives an additional support
that our calculation of $\sigma_R$ is valid 
at the wavelength of 337.1~nm:
no unexpected phenomena (as resonant
absorptions) happened to the nitrogen laser photons
in nitrogen gas. 

The differential cross-section, $d\sigma_R/d\Omega$, for
diatomic molecules such as N$_2$ is modified by a small amount
from the equation~(\ref{eq:dsdth}),
which we used for estimating the number of Rayleigh-scattered
photons entering the PMT (Section~\ref{PhotonAcc}).
This modification
factor at $\theta = 90^\circ$ is $2(1+\rho_0)/(2+\rho_0)$, or 1.011
using $\rho_0 = 0.022$ for the depolarization ratio
of N$_2$ gas induced by the incident light of wavelength 337.1~nm.
For monoatomic molecules such as argon, the depolarization ratio
is zero and $d\sigma_R/d\Omega$ is calculated by
equation~(\ref{eq:dsdth}).
For nitrogen gas, we observed the depolarization effect
in the CRAYS setup as described in Section~\ref{sec:performance}.
We assign a systematic uncertainty of
+1.1\% for $d\sigma_R/d\Omega$.

We observed an elliptical polarization of 4\% for the
incident laser beam with its polarization axis pointing
$37^\circ$ away from the vertical-upward
direction (Section~\ref{sec:performance},
Figure~\ref{fig:polarization}).
Rayleigh scattering of linearly polarized (100\%)
laser beam at $\theta$ = 90$^\circ$ modifies the cross-section by
a factor of
$2(1-{\rm cos}^2\alpha)$, where $\alpha$ is the
rotation angle of the scattered photon measured from
the direction of the polarization \cite{Miles}.
The $\alpha$ is 53$^\circ$ for the CRAYS setup whereas 
$\alpha = 45^\circ$ corresponds to zero correction on the cross-section.
The observed polarization of 4\%
gives a correction factor of $1.000 \pm 0.014$
on the cross-section, corresponding to $\alpha = 45\pm10^\circ$.
We assign a systematic uncertainty
of 1.4\% for the polarization effect.

In summary, for the systematic uncertainty of Rayleigh scattering
cross-section, we have $\pm1.0,~+1.1,~\pm1.4 \%$ from
$\sigma_R$, $d\sigma_R/d\Omega$ and the polarization.
We evaluate a total systematic uncertainty of 2.8\%, taking
a quadratic sum for two $\pm$ uncertainties and adding +1.1\%
uncertainty linearly.

The molecular density of the scatterer gas is calculated from the
temperature (T) and the pressure (P) of the gas inside the CRAYS chamber.
We evaluate an error of 1.3\% for the molecular density calculation,
consisting of the absolute calibration of the mercury barometer (0.5\%), 
the stability of the pressure gauge calibration (1.0\%),
and the difference of the room temperature 
and the gas temperature in the scattering chamber
(maximum 2$^\circ$C corresponding to 0.7\%).

The largest contribution to the systematic uncertainty comes
from the absolute calibration of the energy 
probe \cite{ProbeAccuracy}. The manufacturer calibrated 
the probes with an absolute accuracy of 5\% 
using NIST traceable standards. 
We used two Rjp-465 probes and
the results were
well within the quoted accuracy. 
The second largest contribution comes from
the acceptance calculation, which is dominated by the 
measurement accuracy of the slit 
size (38.9$\pm$0.5~mm) and the distance from the laser
beam line to the PMT mask (312$\pm$3~mm) including the
inaccuracy of the laser beam position in the scattering chamber.
We estimated a total uncertainty of the acceptance to be
3.0\%. 

The uncertainty of $\Sigma_{\rm ADC}$ is estimated as
2.0 \% from the signal integration and 1.9 \% from
the background noise contribution. 
The uncertainty of signal integration (2\%)
is determined from the change of $\Sigma_{\rm ADC}$ by using
a different method of estimating the pedestal level,
and by using different signal integration intervals. 
The uncertainty of background and noise subtraction (1.9\%) is taken
from the remaining $\Sigma_{\rm ADC}$ for
the zero chamber pressure run. It is a conservative estimate
because the amount of the background was stable throughout the
calibration, and its contribution was actually subtracted
in the data analysis.
An uncertainty of 1.0\% was estimated for the geomagnetic effect 
from the change of $\Sigma_{\rm ADC}$ for the YAP run
taken in different azimuthal orientations. 

All added in quadrature, we determined that 
the total systematic uncertainty of the CRAYS calibration
is 7.2\%.

\begin{table}[tb]
\caption{Systematic uncertainties of the CRAYS calibration.}
\label{tb:sys_error}
\begin{center}
\begin{tabular}{|l|r|}
\hline
&Error\\\hline
Cross-section 
($\sigma_R$, $d\sigma_R/d\Omega$ and polarization)    &2.8\%\\
Molecular density (T and P)                           &1.3\%\\
Measurement of laser energy                           &5.0\%\\
Geometric aperture calculation                        &3.0\%\\
Signal integration ($\Sigma_{\rm ADC}$)               &2.0\%\\
Background and noise subtraction ($\Sigma_{\rm ADC}$) &1.9\%\\
Effect of geomagnetism                                &1.0\%\\ \hline
Total (quadratic sum of above)                        &7.2\%\\ \hline
\end{tabular}
\end{center}
\end{table}

\subsection{Transport of the Calibrated PMT}
\label{sec:Transport}
\begin{figure}
\begin{center}
\includegraphics[width=\hsize]{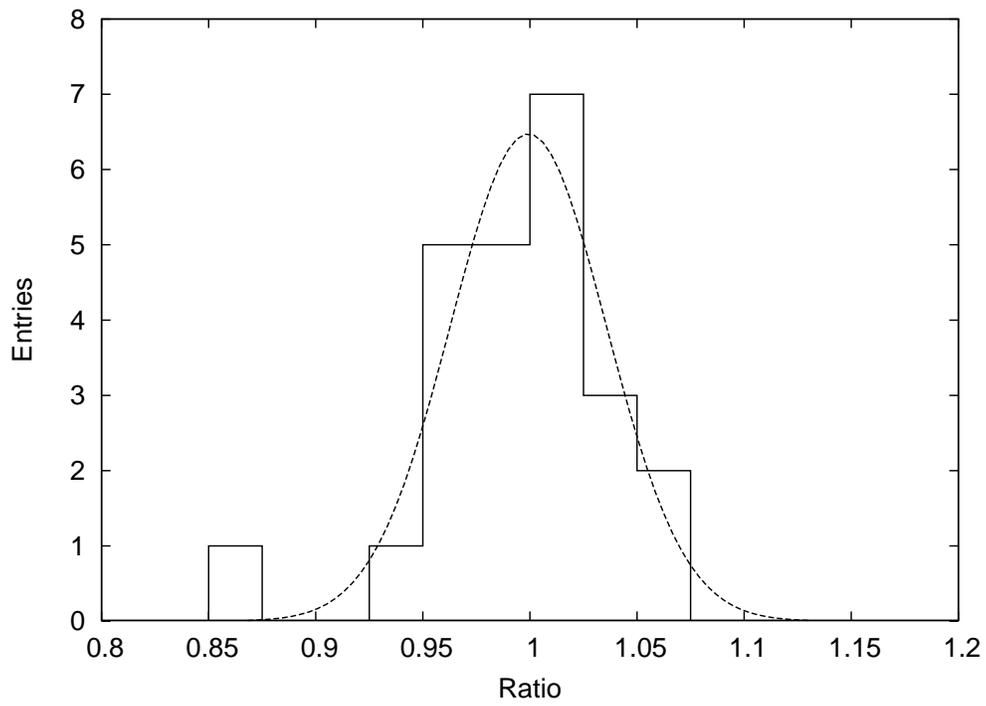}
\end{center}
\caption{Change of the YAP signal from the laboratory calibration 
to the on-site use. A ratio (= on-site/lab.-calib.) is plotted
for 24 Standard PMTs installed in the FD camera.
A fit to the Gaussian
is shown in the dashed line (peak = 0.999, $\sigma$=0.037).
}
\label{fig:YAP_ratio}
\end{figure}

Fifty PMTs with a YAP scintillator were calibrated in January 
2008 in a laboratory of the Institute for Cosmic Ray Research (ICRR),
University of Tokyo\footnote{A second batch of 25 PMTs were calibrated 
in August 2008}. They were then transported to the TA 
site in Utah and installed in the 24 FD cameras in March 2008.
Twenty two cameras had two calibrated PMTs installed and two cameras 
had three calibrated PMTs. 
The same nominal HV setting used in CRAYS calibration 
was applied to 
the Standard PMT installed at the center of the camera, and 
the YAP signal was measured again at the TA site. 
The signal obtained at the site was compared
with that measured during the calibration
after correcting the temperature difference,
25$^\circ$C during the calibration and
$\sim$10$^\circ$C at the TA site,
using the temperature behavior
of the YAP signal previously measured
in the laboratory \cite{temp_dep}. 
The result is
plotted in Figure~\ref{fig:YAP_ratio} as the ratio of the 
two YAP measurements.
Only one PMT showed a large deviation of 0.85,
which is attributable to a change of the YAP light output
\footnote{
Another calibrated PMT installed in the same camera
had the ratio of 1.007, and the gain difference of 
these 2 PMTs was 2.3\% as measured by a xenon flasher run.
}.
The distribution in Figure~\ref{fig:YAP_ratio},
excluding the outlier point (0.85),
is fitted by a Gaussian
with a mean of 0.999 and a standard deviation of 0.037.
The mean value of 0.999
indicates the stability of the PMT gain
from the laboratory calibration to the on-site installation.  
The spread of 3.7\% includes all the following uncertainties
and differences in the measurement: applied HVs,
electronics sensitivities, temperature corrections,
geomagnetic effects in Japan and Utah, 
and possible drifts
of YAP light output and PMT gain during the transport.

\section{Summary}
\label{sec:Summary}
Photometric calibration of the new fluorescence telescope of the TA
was carried out using CRAYS. 
Rayleigh scattering of nitrogen laser 
beam was used for CRAYS to produce a short and uniform UV light flash of known 
intensity on the PMT's photo-sensitive window.
The Standard PMT for each FD camera was calibrated with 
an absolute accuracy of 7.2\% via CRAYS in the laboratory.
An additional uncertainty was introduced by the transport of the calibrated
PMTs from CRAYS to the experimental site in Utah. It is estimated to be 
3.7\% using the YAP pulser.

\section*{Acknowledgements}
We wish to thank the members of the Telescope Array (TA)
collaboration for making it possible (and necessary)
to do this study. 
The TA experiment is supported 
by the Japan Society for the Promotion of Science via
Grants-in-Aid for Scientific Research 
on Specially Promoted Research (21000002) 
``Extreme Phenomena in the Universe Explored by 
Highest Energy Cosmic Rays,'' 
and the Inter-University Research Program 
of the Institute for Cosmic Ray 
Research; by the U.S. National Science Foundation
awards PHY-0307098, 
PHY-0601915, PHY-0703893, PHY-0758342,
PHY-0848320 (Utah), and 
PHY-0649681 (Rutgers); by the National Research 
Foundation of Korea 
(2006-0050031, 2007-0056005, 2007-0093860,
2010-0011378, 2010-0028071, R32-10130);
by the Russian Academy of Sciences (RFBR grants 
10-02-01406a and 11-02-01528a (INR)),
IISN project No. 4.4509.10; and the 
Belgian Science Policy under IUAP VI/11 (ULB).
We also wish to thank the people and officials of Millard County,
Utah, for their wholehearted support. 
We gratefully acknowledge the contributions 
from the technical staffs of our
home institutions and the Center 
for High Performance Computing (CHPC), University of Utah.

\clearpage\newpage

\clearpage\newpage

\clearpage\newpage

\clearpage\newpage

\clearpage\newpage

\clearpage\newpage

\clearpage\newpage

\clearpage\newpage

\clearpage\newpage

\clearpage\newpage

\clearpage\newpage

\clearpage\newpage

\clearpage\newpage

\clearpage\newpage

\clearpage\newpage

\end{document}